\newcommand{\be}{\begin{equation}}
\newcommand{\ee}{\end{equation}}

\documentclass[a4paper]{article}
\usepackage{latexcad}
\begin{document}
\begin{center}
\vskip .2in

{\large \bf Time and Events}
\vskip .50in
\vskip .2in

Ph.~Blanchard\footnote{Faculty of Physics and BiBoS,
University of Bielefeld,
Universit\"atstr. 25,
D-33615 Bielefeld,
e-mail: blanchard@physik.uni-bielefeld.de}
\ and\ A.~Jadczyk\footnote{Institute of Theoretical Physics,
University of Wroc{\l}aw,
Pl. Maxa Borna 9,
PL-50 204 Wroc{\l}aw,
e-mail: ajad@ift.uni.wroc.pl}

\end{center}

\vskip .2in

\begin{abstract}
Time plays a special role in Standard Quantum Theory. The concept of
time observable causes many controversies there. In Event Enhanced
Quantum Theory (in short: EEQT) Schr\"odinger's {\em differential
equation} is replaced by a {\em piecewise deterministic algorithm}
that describes also the {\em timing of events}. This allows us to
revisit the problem of time of arrival in quantum theory.
\end{abstract}
\newpage
\section{Introduction}
EEQT was invented to answer John Bell's concerns about the status
of the measurement problem in quantum theory (Bell 1989,1990). EEQT's main thesis
is best summarized in the following statement:
\begin{center}
{\em NOT ALL IS QUANTUM}
\end{center}
Indeed, a pure quantum world would be dead. There would be no
events; nothing would ever happen. There are no dynamics in pure
quantum theory that serve to explain how potentialities become
actualities. And we do know that the world is not dead. We know events
do happen, and they do it in finite time. This means that pure quantum
theory is inadequate. John Bell realized this fact and at first he
sought a solution in hidden variable theories (Bell 1987a). Rudolf Haag (Haag 1990a,1990b,1996a,1996b) takes a similar position; he calls it an "evolutionary picture." EEQT is motivated
by the same concerns but has taken a slightly different perspective.
What EEQT has in common with hidden variable theories (as well as John
Bell's "beables" (Bell 1987b)) is the realization that
\begin{center}
{\em THERE IS A CLASSICAL PART OF THE UNIVERSE}
\end{center}
and this part can evolve. "We" (IGUS-es) belong partly to this
classical world. 
Once the existence of the classical part is accepted, then {\em
events} can be defined as changes of state of this classical part - cf. Fig.\ref{fig:tset}
EEQT is the only theory that we are aware of that can precisely define the two
concepts:
\begin{itemize}
\item EXPERIMENT
\item MEASUREMENT
\end{itemize}
thus complying with the demands set by John Bell in (Bell 1989,1990).
We define (cf. Jadczyk 1994))
\begin{itemize}
\item {\bf Experiment}: completely positive one--parameter family
of maps of ${\cal A}_{tot}$
\item {\bf Measurement}: very special kind of experiment
\end{itemize}
Moreover, EEQT is the only theory where there is one--to--one correspondence
between linear Liouville equation for ensembles and individual \underline{algorithm}
for generation of events. It is to be noted that, although EEQT does not involve hidden
variables, it does seek for deeper than just statistical descriptions. Namely it asks
the following question:
\begin{quote}
\underline{HOW} NATURE CREATES THE UNIQUE WORLD AROUND US AND OUR OWN UNIQUE
PERCEPTIONS?
\end{quote}
In other words EEQT seeks for knowledge by going beyond pure descriptive
orthodox interpretation.\footnote{We would like to quote at this place this, simple
but deep, wisdom {\em Knowledge protects, ignorance endangers.}} In agreement with
Rudolf Haag, we are taking an evolutionary view of Nature. This means that the Future
does not yet exist; it is being continuously created, and this creation is
marked by events. But \underline{how} does this process of creation
proceed? This is what we want to know. A hundred years ago the answer would have been:
"by solving differential equations." But today, after taking lessons in
relativity and quantum theory, after the computer metaphor has permeated so
many areas of our lives, we propose to seek an answer to the question of
"how?" in terms of an ALGORITHM. Thus we set up the hypothesis: {\em Nature
is using a certain algorithm that is yet to be discovered}. Quantum theory
tells us clearly: this algorithm is non-local. Relativity adds to this:
non-local in space implies non-local in time. Thus we have to be prepared
to meet acausalities in the individual chains of events even if they
average out in big statistical samples. EEQT can be thought of as
one step in this direction. It proposes its piecewise deterministic process
(PDP of Ref. (Blanchard and Jadczyk, 1995)) as the algorithm for generating a sample
history of an individual system. This algorithm should be thought of
as a fundamental one, more basic than the Master Equation which
follows from it after taking a statistical average over different possible
individual histories. In EEQT it still holds that there is one--to--one
correspondence between PDP and the Master Equation, and it is easy to think
of an evident and unavoidable generalization of PDP, when feedback is
included, which will go beyond the linear Master Equation and thus
beyond linear Quantum Theory. Work in this direction is in progress.

According to the philosophy of EEQT, the quantum state vector is an
auxiliary variable which is not directly observable, even in part.
It is a kind of a hidden variable. But, according to EEQT, there
\underline{are} directly observable quantities - and they form the
${\cal A}_{clas}$ part of ${\cal A}_{tot}$. EEQT does not assume
standard quantum mechanical postulates about results of measurements
and their probabilities. All must be derived from the dissipative
experiment dynamics by observing the events at the classical level
i.e. by carring out continuous observation of the state of
${\cal A}_{tot}$.

An \underline{event} is thus a fundamental concept in EEQT and there are
two primitive event characteristics that the algorithm for event generation
must provide: the "when" and the "which," and indeed PDP is a piecewise
deterministic algorithm with {\em two} "roulette" runs for generating each
particular event. First the roulette is run to generate the {\em time of event}.
Only then, after the timing has been decided, there is a second roulette run
that decides, according to the probabilities of the moment, {\em which} of
the possible events is selected to occur. Then, once these two choices have been
made, the selected event happens and is accompanied by an appropriate quantum
jump of the wave function. After that, continuous evolution of possibilities
starts again, roulette wheels are set into motion, and the countdown begins for
the next event.
\section{Event Generating Algorithm}
No event can ever happen unless a given quantum system is coupled to
a classical system. In fact, the Reader should be warned here that this
statement is not even precise. A precise statement would be: "no event can
happen to a system unless it contains a classical subsytem." In many cases,
however, the total system can be considered a direct product of a pure quantum
system and a classical one. If we restrict ourselves to such a case, then the
simplest nontrivial event generator is a "fuzzy property detector" defined as
follows. Let $Q$ be a pure quantum system whose (uncoupled) dynamics are described by a self--adjoint Hamiltonian $H$ acting on a Hilbert space ${\cal H}$. A fuzzy
property detector is then characterized by a positive operator  $F$ acting on ${\cal H}$. In the
limit of a "sharp" property we would have $F^2=\sqrt{\kappa} F$, where $\kappa$ is a numerical
coupling constant (of physical dimension $t^{-1}$). That is the property becomes sharp
for $F$ proportional to an orthogonal projection.\\
According to a general theory described in (Blanchard and Jadczyk, 1995), a property detector
is a two-state classical device, with states denoted $0$ and $1$ and characterized by
the transition operators (using the notation of (Blanchard and Jadczyk, 1995)): $g_{01}=0$, $g_{10}=F$.
The Master Equation describing continuous time evolution of statistical states of the
quantum system coupled to the detector reads:
\begin{eqnarray}
{\dot \rho}_0(t)&=&-i[H_0,\rho_0(t)]+F \rho_1 F
\nonumber\\
{\dot \rho}_1(t)&=&-i[H_1,\rho_1]-{1\over2}
\{F^2,\rho_1\}.
\end{eqnarray}
Suppose at $t=0$ the detector is off, that is in the state denoted by $0$, and the particle state is
$\psi (0)$, with $\Vert \psi (0)\Vert=1.$
Then, according to the event generating algorithm
described heuristically in the previous section, the probability $P(t)$ of detection, that is of a change of state
of the classical device, during
time interval $(0,t)$ is equal to $1-\Vert K(t)\; \psi (0)\Vert^2 $,
where $K(t)=\exp (-iH_0 t-{\Lambda\over2}t)$, where $\Lambda=F^2$.
It then follows that the probability
that the detector will be triggered out in the time interval (t,t+dt),
provided it was not triggered yet, is $p(t)dt$, where $p(t)$ is
given by
\be
p(t)={d\over{dt}}P(t)=<K(t)\psi_0,\Lambda K(t)\psi_0>.
\label{eq:pst}
\ee
Let us consider the case of a maximally sharp measurement. In this case
we would take $\Lambda=|a><a|$, where $|a>$ is some Hilbert space vector.
It is not assumed
to be normalized; in fact its norm stands for the strength of the coupling
(note that $<a|a>$ must have physical dimension $t^{-1}$).
From this formula it can be easily shown (Blanchard and Jadczyk, 1996) that
$p(t)=\vert \phi(t)\vert^2$, where the Laplace transform ${\tilde\phi}(z)$ of the (complex)
amplitude $\phi(t)$ is given by the formula
\be
{\tilde\phi}={2<a|{\tilde K_0}|\psi_0>
\over{2+{<a|{\tilde K}_0|a>}}}
\label{eq:atp}
\ee
where $K_0(t)=\exp (-iH_0 t)$.
\section{Time of Arrival}

Let us consider a particular case of {\sl time of arrival}\ (cf.
Muga et al. 1995) for a recent discussion). Thus we take $|a>$ to denote
a position eigenstate localized at the point $a$, that is 
$<x|a>=\sqrt{{\kappa}}\delta(x-a)$, $\kappa$
being a coupling constant representing efficiency of the detector. For the
Laplace transform ${\tilde\phi}$ of the probability amplitude we obtain then
\be
{\tilde \phi}=
{{2\sqrt{\kappa}}\over{2+{\kappa}{\tilde K}_0(a,a)}}{\tilde \psi}_0(a)
\label{eq:ak}
\ee
where
${\tilde \psi}_0$ stands for the Laplace transform of $K_0(t)\psi_0$.\\
Let us now specialize to the case of free Schr\"odinger's particle on a line. 
We will study response of the point counter to a Gaussian wave packet whose initial shape
 at $t=0$ is given by:
\be
\psi_0(x)=
{
1
\over
{
(2\pi )^{1/4}\eta^{1/2}
}
}\exp 
\left(
{{ -(x-x_0)^2}\over{4\eta^2}}
+2ik(x-x_0)
\right) .
\ee
In the following it will be convenient to use dimensionless variables
for measuring space, time and the strength of the coupling:
\be
\xi={x\over 2\eta},\qquad \tau={{\hbar t}\over{2m\eta^2}},\qquad
\alpha={{m\eta\kappa}\over\hbar}
\ee
We denote 
\be
\xi_0=x_0/2\eta ,\; \xi_a=a/2\eta ,\; v=2\eta k,  
\ee

\be
u_{\pm}=i\sqrt{-iz}\, \pm(v-id) , \qquad d=\xi_0-\xi_a .
\ee

\begin{figure}[	t]
\epsfysize=8cm
\epsffile{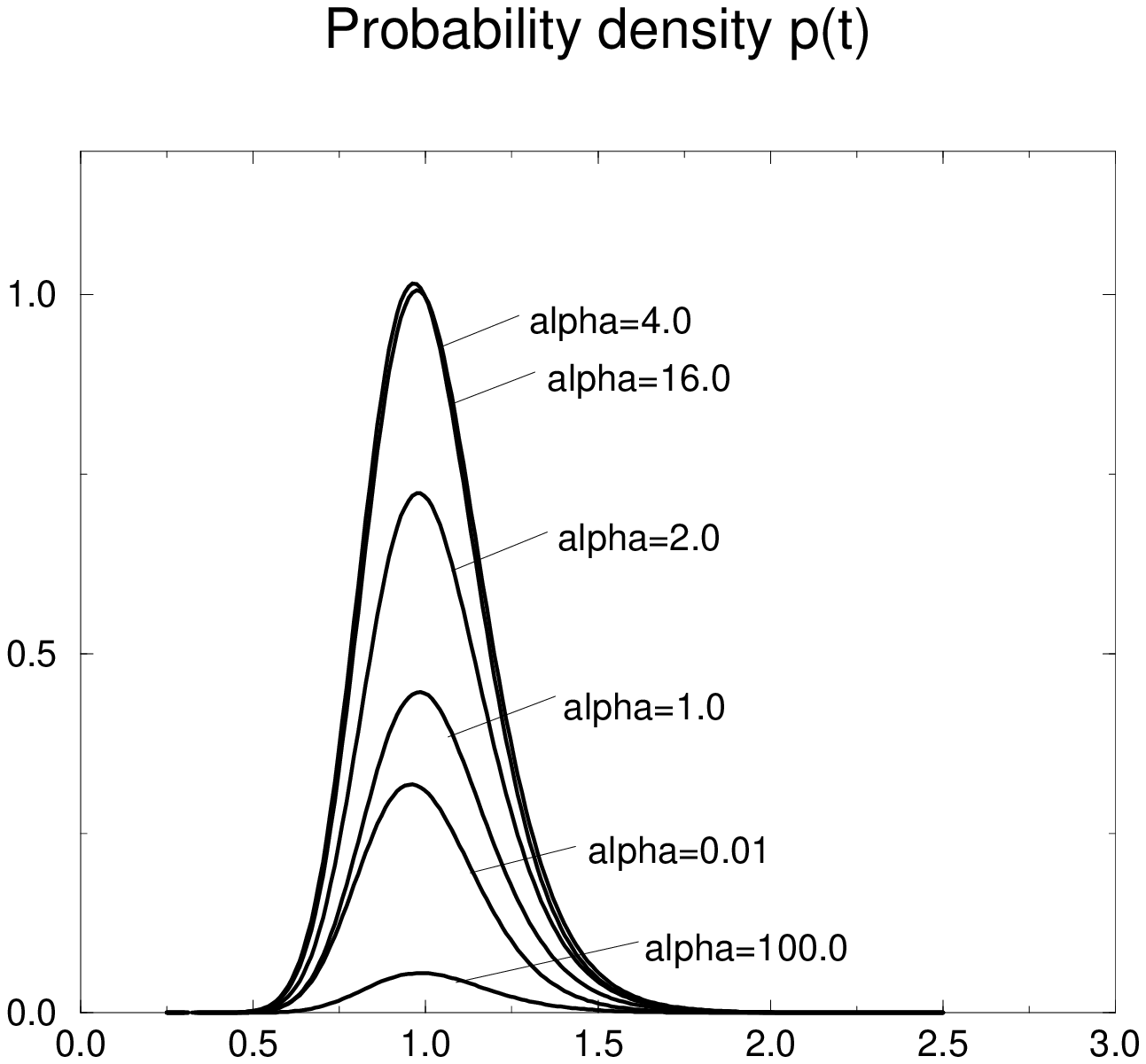}
\caption{Probability density of time of arrival for a Dirac's delta counter
placed at $x=0$, coupling constant alpha. The incoming wave packet starts at $t=0$,
$x=-4$, with velocity $v=4$}
\label{fig:fig2}
\end{figure}

The amplitude ${\tilde\phi}$ of Eq. (\ref{eq:ak}), when rendered
dimensionless reads then
\be
{\tilde\phi}(z)=(2\pi)^{1/4}\alpha^{1/2}e^{-d^2-2ivd}\, 
\frac{\mbox{w}(u_+) +\mbox{w}(u_-)}
{2\sqrt{iz}\, +\alpha}
\ee
with the function $w(u)$ defined by
\be
\mbox{w}(u)=e^{-u^2}\, \mbox{erfc}(-iu)
\ee
The time of arrival probability curves of the counter for several values of
the coupling constant are shown in Fig.\ref{fig:fig2}. The incoming wave packet starts at 
$t=0$, $x=-4$, with velocity $v=4 .$ It is seen from the plot that the average
time at which the counter, placed at $x=0$, is triggered is about 
one time unit, independently of the value of the coupling constant. 
This numerical example shows that
our model of a counter can be used for measurements of time of
arrival. It is to be noticed that the shape of the response curve is 
almost insensitive to the value of the coupling constant. It is also important to
notice that in general the probability $P(\infty)=\int_0^\infty p(\tau)d\tau$
that the particle will be detected at all is less than $1.$ In fact, for a pointlike
counter as above, the numerical maximum is $<0.73$ - cf. (Blanchard and Jadczyk, 1996). 
For this
reason (i.e. because of the need of normalization) {\sl time of arrival observable is
not represented by a linear operator.}
\section*{Acknowledgment(s)} 
One of us (A.J.) would like to thank A. von Humboldt Foundation for
the support. He also thanks to L. Knight for reading the manuscript.\\
{\bf REFERENCES}\\
Bell,  J. (1989) \lq Towards an exact quantum mechanics\rq,  in
{\sl Themes in Contemporary Physics II.  Essays in honor of Julian
Schwinger's 70th birthday},  Deser,  S. ,  and Finkelstein,  R. J.  Ed. ,
World Scientific,  Singapore \\
Bell,  J. (1990) \lq Against measurement\rq, in
{\sl Sixty-Two Years of Uncertainty. Historical, Philosophical and
Physical Inquiries into the Foundations of Quantum Mechanics}, Proceedings
of a NATO Advanced Study Institute, August 5-15, Erice, Ed. Arthur I. Miller,
NATO ASI Series B vol. 226 , Plenum Press, New York \\
Bell, J. (1987a) \lq On the impossible pilot wave\rq , in {\sl
Speakable and unspeakable in quantum mechanics}, Cambridge University Press\\
Bell, J. (1987b)\lq Beables for quantum theory\rq, opus cited (Bell 1987a)\\
Blanchard, Ph., and A. Jadczyk, (1995) \lq
Event--Enhanced--Quantum Theory and Piecewise Deterministic Dynamics\rq,
{\em Ann. der Physik} {\bf 4} 583--599; see also the short version: \lq Event and Piecewise
Deterministic Dynamics in Event--Enhanced Quantum Theory\rq\ , {\em Phys.Lett.}
{\bf  A 203} 260--266\\
Blanchard, Ph., Jadczyk, A. (1996) \lq Time of Events in Quantum
Theory\rq , {\sl Helv. Phys. Acta} {\bf 69} 613--635\\
Haag, R. (1990a) \lq Irreversibility introduced on a fundamental level\rq ,
{\sl Commun. Math. Phys.} {\bf 123}  245-251\\
Haag, R. (1990b) \lq Thought of the Synthesis of Quantum Physics and
General Relativity and the Role of Space--time\rq , {\sl Nucl. Phys.} {\bf B18}
135--140\\
Haag, R. (1995) \lq An Evolutionary Picture for Quantum Physics\rq ,
{\sl Commun. Math. Phys.} {\bf 180}  733-743\\
Haag, R. (1996) {\sl Local Quantum Physics}\ 2nd rev. and enlarged
ed. , Ch. VII: {\em Principles and lessons of Quantum Physics. A Review of 
Interpretations, Mathematical Formalism and Perspective.}\\
Jadczyk, A. (1995) \lq On Quantum Jumps, Events and Spontaneous
Localization Models\rq, {\em Found. Phys.} {\bf 25}  743--762\\
Muga, J.G., Brouard, S., and Macias, D. (1995) \lq Time
of Arrival in Quantum Mechanics\rq , {\em Ann. Phys.} {\bf 240}
351--366\\
\newpage
\vskip10cm
\placedrawing{berlin1.lp}{}{fig:tset}
\pagebreak

\end{document}